\documentclass[sigconf,screen,nonacm]{acmart}
\AtBeginDocument{%
  \providecommand\BibTeX{{%
    \normalfont B\kern-0.5em{\scshape i\kern-0.25em b}\kern-0.8em\TeX}}}

\setcopyright{acmcopyright}
\copyrightyear{2022}
\acmYear{2022}
\acmDOI{XX.XXXX/XXXXXXX.XXXXXXXX}

\acmConference[Web Science '22]{Web Science '22: ACM Web Science Conference}{June 26--29, 2022}{Barcelona, Spain}
\acmBooktitle{Web Science '22: ACM Web Science Conference, June 26--29, 2022, Barcelona, Spain}
\acmPrice{15.00}
\acmISBN{978-1-4503-XXXX-X/18/06}

\begin{document}

\title{Bots don't Vote, but They Surely Bother!} 
\subtitle{A Study of Anomalous Accounts in a National Referendum}

\author{Eduardo Graells-Garrido}
\affiliation{%
  \institution{Data Science Institute, Universidad del Desarrollo}
  \city{Santiago}
  \country{Chile}
}
\email{egraells@udd.cl}
\author{Ricardo Baeza-Yates}
\affiliation{%
  \institution{Institute for Experiential AI, Northeastern University}
  \state{California}
  \country{USA}
}
\email{rbaeza@acm.org}

\renewcommand{\shortauthors}{Graells-Garrido and Baeza-Yates}
\begin{abstract}
The Web contains several social media platforms for discussion, exchange of ideas, and content publishing. These platforms are used by people, but also by distributed agents known as bots. 
Although bots have existed for decades, with many of them being benevolent, their influence in propagating and generating deceptive information in the last years has increased. 
Here we present a characterization of the discussion on Twitter about the 2020 Chilean constitutional referendum. 
The characterization uses a profile-oriented analysis that enables the isolation of anomalous content using machine learning. As result, we obtain a characterization that matches national vote turnout, and we measure how anomalous accounts (some of which are automated bots) produce content and interact promoting (false) information. 
\end{abstract}

\ccsdesc[500]{Information systems~Social networks}

\keywords{Social networks, bot detection, political polarization, stance classification.}

\maketitle

\section{Introduction}

Social media platforms have acquired a crucial role in meaning-making processes within communities \cite{kavada2020creating}. In the context of social changes and worldwide events, such processes have acquired more importance than ever. As technology evolves, the ``social'' has become more than just people: social platforms provide a myriad of services ranging from news, health, business, games, among others. The entities in these platforms are people, but also companies, political parties, and media sources of all sizes and credibility. Yet, not all accounts that pretend to be people are actual persons. Some of them are automated accounts. Although sometimes bots are benevolent~\cite{aiello2012people}, the last several years the focus of bots has been deceiving people by manipulating and amplifying social media content. This situation has promoted methods to detect and characterize bots~\cite{ICWSM1715587}, as well as to understand their role in social interactions mediated by these platforms~\cite{shao2018spread}.

In this paper we study the political discussion around the Chilean Constitutional Referendum, held in October 25th, 2020. This event was one of the consequences of the fiercest social outburst in the last decades~\cite{somma2021no}. It started on October 18, 2019, and it is considered an important event that has impacted Chile's well-being, due to a ``perfect storm'' of situations, including the recent pandemic~\cite{morales2021chile}. 
One of the main demands of the social movements involved was a referendum to draft a new constitution for the country, because the current constitution was drafted during Pinochet's dictatorship. Thus, the plebiscite enticed strong and polarizing discussions on social media, particularly in the micro-blogging platform Twitter. Being publicly accessible, the trending topics of Twitter are part of everyday conversations and media reports. 
Given how social media can shape people's perception, and how this perception can be tied to voting turnout, here we aim to understand the role of bots in the discussion. Mainly, we focused on the volume of content published by bots, their potential synchronization, and their political leaning.

We applied an existing methodology for stance detection (not referenced for anonymity), 
which enabled us to classify Twitter accounts into in-favor or against a new constitution. Then, we applied an existing anomaly detection method, Isolation Forest~\cite{liu2008isolation,liu2012isolation}, to quantify how anomalous was each account with respect to their behavior in the platform. We interpreted the global patterns of anomalous behavior, and then established a criteria to define a bot. As result, we observed that the stance classification produced results aligned with the election turnout; that the fraction of bots is small (0.66\%) but their impact is much larger; and that, in terms of interaction and information diffusion, there are bot communities in both sides of the political spectrum, yet the larger communities were right-leaning, against the drafting of a new constitution.

\section{Data}

\begin{figure}
    \includegraphics[width=0.95\linewidth]{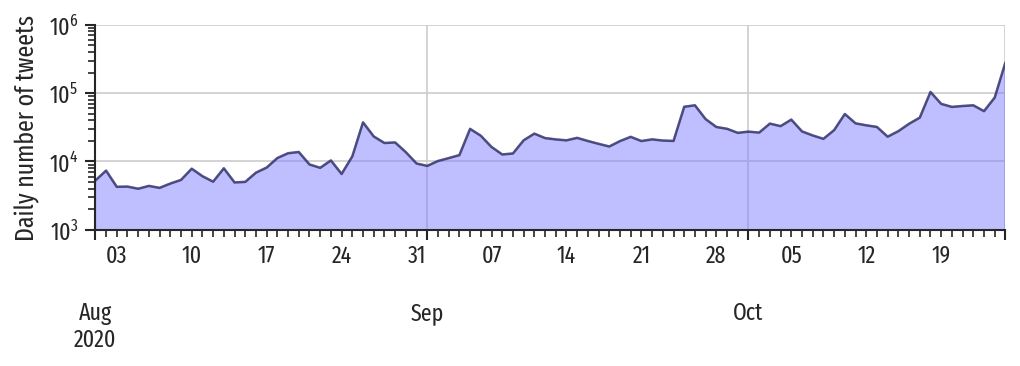}
    \caption{Weekly volume of content in the data set.}
    \label{fig:tweet_volume}
\end{figure}

We connected to the Twitter Streaming API using a system designed to crawl Chilean tweets. 
The query parameters were keywords related to mainstream political discussion in Chile, including keywords related to the two stances of the referendum: to approve (\emph{Apruebo} in Spanish) the drafting of a new constitution, or to reject it (\emph{Rechazo} in Spanish). We studied the period between August 1st, 2020, and October 25th, 2020. In total, we obtained 2.3M tweets from 251K users (see Figure~\ref{fig:tweet_volume}) after a cleaning process. This represents about 10\% of all Twitter users in Chile\footnote{\url{https://datareportal.com/reports/digital-2020-chile}, p. 38.} and about 1.3\% of the Chilean population at that time. Of those tweets, 32\% were retweets, 6\% were quotes of other tweets, and 9\% were replies to other tweets. 
  
\section{Predicting Stance}
Given the size of the discussion under analysis, manually labeling the user profiles into the two stances \{\emph{apruebo}, \emph{rechazo}\} is expensive and impractical. In view of this difficulty, we predicted stances using a classifier trained on a labeled subset of the data set.
This subset is labeled automatically from a list of seed patterns and keywords for each stance, as they are an effective mechanism to predict the community a user belongs to~\cite{bryden2013word}.

\begin{figure}
    \centering
    \includegraphics[width=0.9\linewidth]{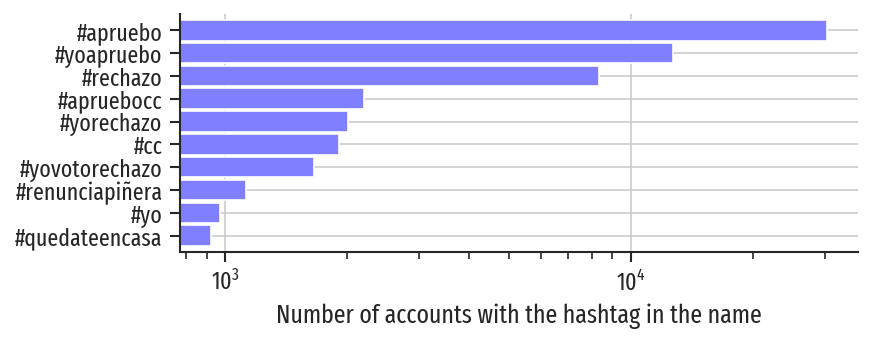}
    \caption{Frequency of top hashtags found in full names within profiles.}
    \label{fig:profile_hashtags}
\end{figure}

To identify seeds, we explored the data set to seek for terms that could be mapped to the \emph{apruebo} and \emph{rechazo} stances. This included hashtags (\emph{\#apruebo}, \emph{\#yoapruebo} --I approve--, \emph{\#votoapruebo} --I vote approve--; and their counterparts). We noticed that a relevant fraction of users self-reported their stances in the full name section of their profiles by including hashtags (see Figure~\ref{fig:profile_hashtags}). 
The seed terms are not necessarily frequent, but they are discriminating, {\em i.e.}, it is likely that someone in its corresponding category would use the term, and not from the other. The list is built iteratively in the sense of running the first steps from this section up to the classification step, and then exploring the usage of discriminating terms by accounts in each group to look for other potential seeds. Additionally, when we observed an account that was remarkably associated with a stance but was classified as the other, we manually labeled that account. In total, we manually labeled just 8 \emph{apruebo} accounts, and 50 \emph{rechazo} accounts.

Next, we propagated the user labels from the previous step to the rest of the data set. We used the XGBoost classifier that trains decision trees using gradient boosting~\cite{chen2016xgboost}. The~input feature matrix is the concatenation of several matrices:
\begin{itemize}
\item An account-term matrix, that encodes the number of times each account has used each term.
\item A profile-term matrix, analogous to the previous one, but~this time for the terms contained in the full name and biographical self-description of each user.
\item A profile-domain matrix, mapping to each user's home page its main domain ({\em e.g.}, twitter.com) and their main top level domain ({\em e.g.}, .com).
\item Three adjacency matrices based on the interactions in the discussion: retweets, replies, and~quotes. 
\item A user-stance interaction matrix for each type of interaction, where each cell contains the number of times the corresponding user has interacted with other users that were pre-labeled with a stance. 
\end{itemize}
We removed terms that were used for labeling from the feature matrix, as they perfectly separate users from both groups and our goal is to classify users who do not use these terms in their content. Then, we trained the classifier using the set of labeled users.  

\begin{figure*}
    \includegraphics[width=0.85\linewidth]{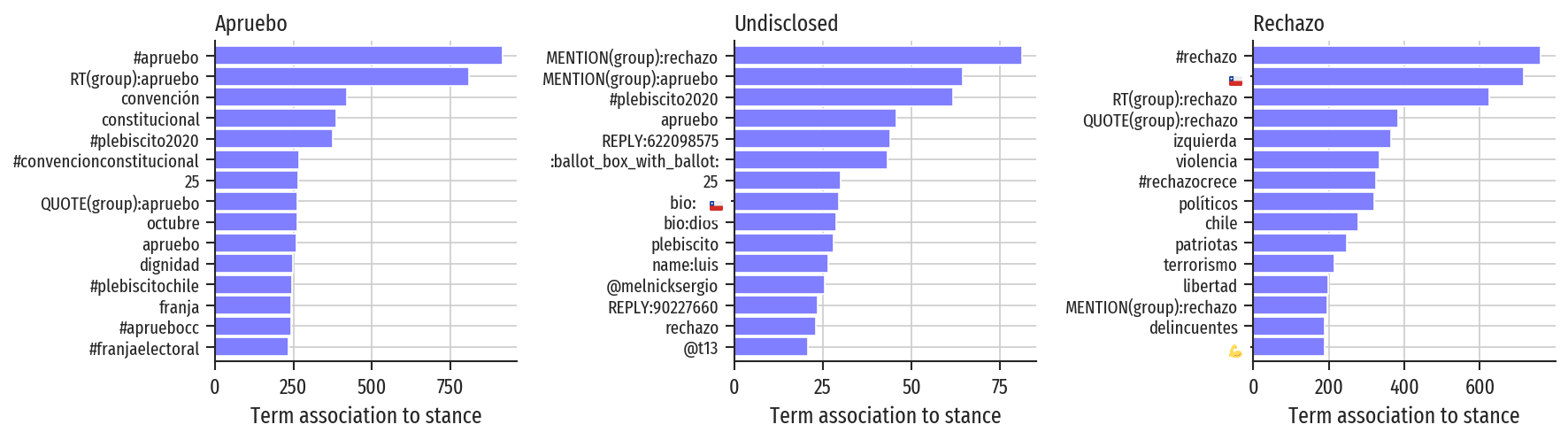}
    \caption{Top terms and features associated with each stance.}
    \label{fig:term_associations}
\end{figure*}

Then, we predicted the stance of the rest of the data set. For a given account $u$, the classifier outputs a value $p_a(u)$ for each stance $a$ that lies in $[0,1]$, corresponding to the fraction of decision trees that vote for the corresponding stance. We applied a small threshold ($p_a(u) \geq 0.55$) to consider predictions with at least a small confidence by the classifier. Those accounts who cannot be classified were marked as \emph{undisclosed}. 

As result, we predicted 81.20\% of accounts in \emph{apruebo}, 17.34\% of accounts in \emph{rechazo}, and 1.46\% as \emph{undisclosed}. This matches well the referendum results, where 78.28\% voted \emph{apruebo}, whereas 21.72\% voted \emph{rechazo}. The distribution is similar, hinting that Twitter is a powerful signal when analyzing national-level events, even when the sample is not representative of the overall country demographics. This was a result obtained even before the culmination of the study, as one week before the election we shared a preliminary prediction using this method with journalists. 

To characterize each stance, we estimated which terms and features were most associated with each stance using the log-odds ratio (see Figure~\ref{fig:term_associations}). As expected, features related to homophily are important, such as retweeting and quoting pre-labeled users. With respect to content, we found that in \emph{apruebo}, words like \emph{dignidad} (dignity) and \emph{pueblo} (people) exhibit its political left-leaning, whereas the Chilean flag emoji, mentions to \emph{libertad} (liberty) and \emph{patriotas} (patriots) exhibit its right-leaning nationalism.

\begin{figure}
    \centering
    \includegraphics[width=\linewidth]{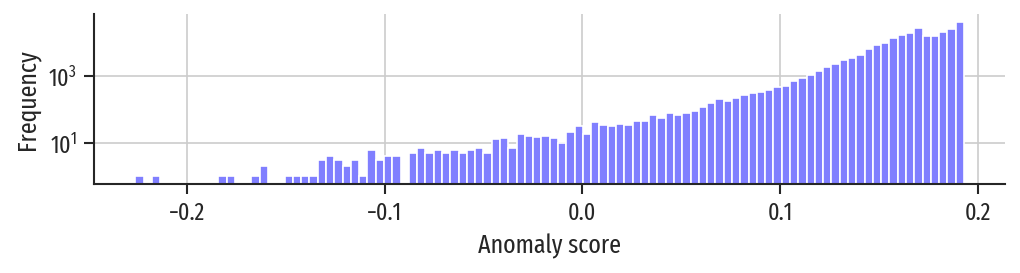}
    \caption{Distribution of anomaly scores.}
    \label{fig:if_scores}
\end{figure}

\section{Quantifying Anomalous Behavior}

We want to know to which extent the discussion was influenced by anomalous accounts. First, we downloaded the largest data set of bots available~\cite{feng2021twibot}, and found that only two accounts from our data set were on it, both wrongly marked as bots: one account was from a legitimate media platform and the other was from a right-wing politician. 
Since we do not have known bot labels for users, and bot classifiers, such as Botometer,\footnote{\url{https://botometer.osome.iu.edu/}.} tend to rely on the full recent content published by accounts (in contrast to our content-based crawling approach), and may not work well in other languages than English~\cite{rauchfleisch2020false}, we implemented a method to quantify the anomaly (or lack thereof) of every account in our study.

We based our work in the Isolation Forest (IF)~\cite{liu2008isolation,liu2012isolation} model. 
IF is an unsupervised anomaly detection that quantifies the distance of a given observation to the rest of the data. The model has two assumptions regarding anomalies: first, they are few; second, they are very different to normal observations. 
As such, anomalies can be succinctly described with respect to the rest of the data set. 
The model does so by building an ensemble of trees, where each tree learns a description of a sample of the data based on binary partitions of its feature space. An anomaly score for an observation is derived from the average path length in all trees. It is equivalent to the number of splittings required to isolate the observation. The greater the score, the less anomalous an observation is. 

\begin{figure}
    \centering
    \includegraphics[width=0.95\linewidth]{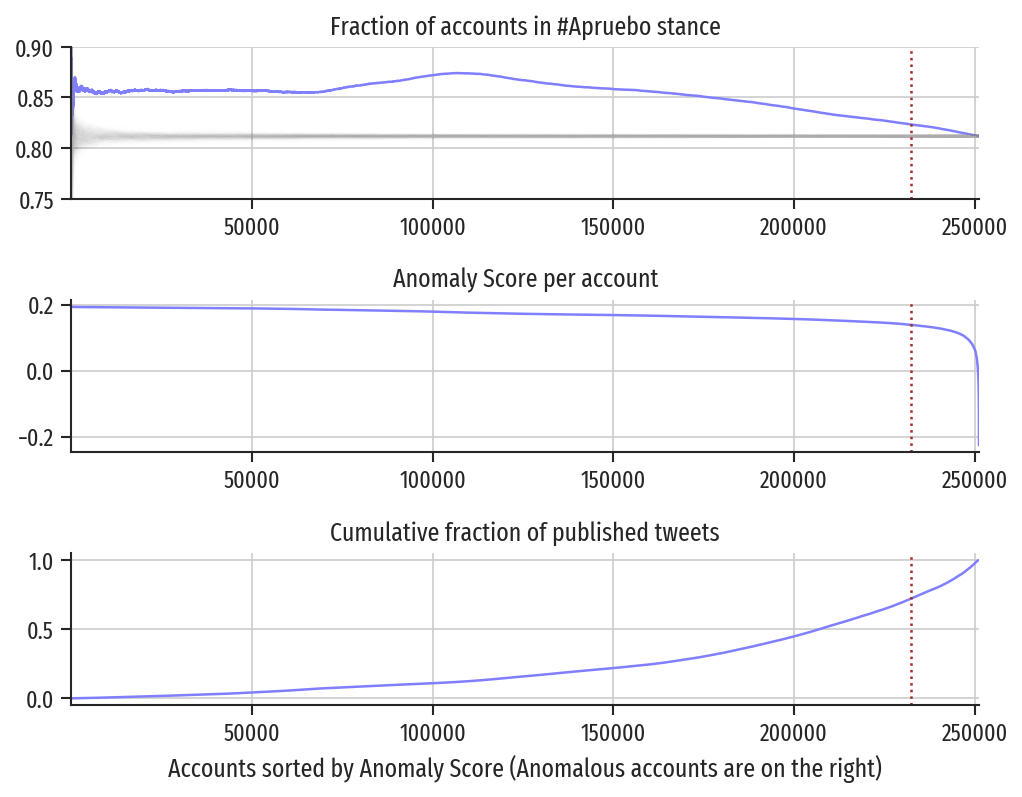}
    \caption{Top: fraction of accounts in \emph{apruebo} (blue line) with respect to the number of accounts over the anomaly threshold (red dotted line), compared with a null model (gray lines). Center: Anomaly scores per account. Bottom: Cumulative fraction of published tweets.}
    \label{fig:scores}
\end{figure}

To define how to estimate the distance between accounts, we built a feature matrix with the following elements per account:
\begin{itemize}
    \item The number of active days in the data set, {\em i.e.}, the number of days with published content by each account.
    \item Relative amount of content: the number of published tweets, retweets, quotes, and replies, log-transformed and divided by the amount of active days.
    \item A daily rhythm, consisting of the total amount of content published divided by the number of active days.
    \item Ratio of friends (followees) over followers.
    \item The number of digits in the account username.
    \item A flag regarding the use of the default profile image.
    \item Whether the account is on a connected component of interactions, and which component.
    \item Account age in days.
    \item Relative global behavior: the log-transformed number of globally published tweets, friends, and followers, divided by the account age.
\end{itemize}

After applying the model, we sorted the accounts with respect to their anomaly scores (see Figure ~\ref{fig:if_scores} for the distribution). In data sets of similar size, it has been determined that around 7.5\% of accounts are bots~\cite{mendoza2020bots}. We looked at the distribution of anomaly scores and the cumulative fraction of published tweets (see Figure ~\ref{fig:scores}, center), and we observed that anomaly scores present relevant values in a smaller proportion of accounts, and that anomalous account tend to publish more than normal ones. It is known that the distribution of published tweets follows a power-law and that few accounts generate most of the content. Those accounts could also be anomalous, but not necessarily bots.

To understand whether there is a relationship between anomaly and political position, we compared the fraction of accounts in \emph{apruebo} at every incremental subset of accounts sorted by anomaly score. We compared this distribution with a null model where the political stance was permuted at random (see Figure~\ref{fig:scores}, top). In the null model, the anomaly score is not correlated with the fraction of accounts in \emph{apruebo}, which is what we expected. However, in the observed distribution, there is a complex relationship between being anomalous and the fraction of accounts in \emph{rechazo}. This result hints that most anomalous activity is associated with right-wing politics. Bots need a significant investment, so this is not surprising either.

\section{Discriminating Bots}

\begin{figure}
    \centering
    \includegraphics[width=\linewidth]{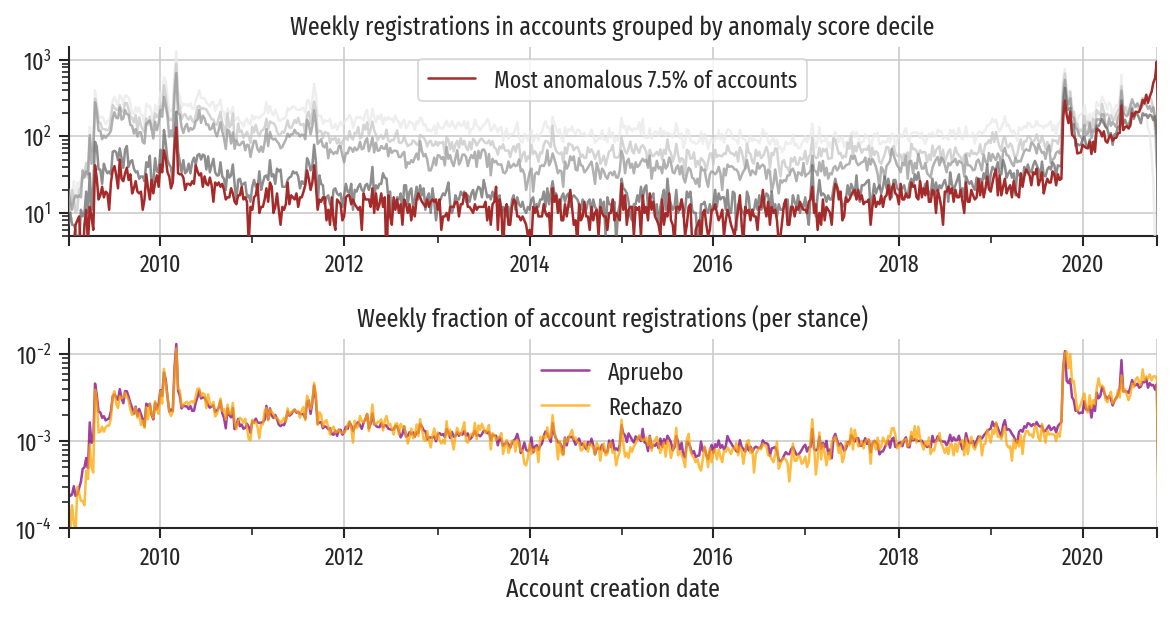}
    \caption{Weekly registrations with respect to anomaly score (top), and stance (bottom, normalized).}
    \label{fig:registrations}
\end{figure}

The anomaly score points accounts that could be bots, but additional criteria is needed to identify them. One element of this criteria is to consider the age of accounts: old accounts may be anomalous with respect to their behavior, such as a high frequency of publications, but this may be a natural behavior of the population. Hence, we separated accounts into five groups, with the first group defined as the 7.5\% most anomalous, and the other four groups defined as the subsequent accounts in evenly-spaced ranges of anomaly score (see Figure~\ref{fig:registrations}). We observe that the most anomalous group tends to be on the lower-bound of registrations per week until one week after the beginning of the study (August 8th, 2020). After that, it became the most active group in terms of registrations. In comparison, the distribution of registrations with respect to stance does not present differences between stances, implying that registration date may help us to point to bots regardless of their political position. 

\begin{figure}
    \centering
    \includegraphics[width=0.95\linewidth]{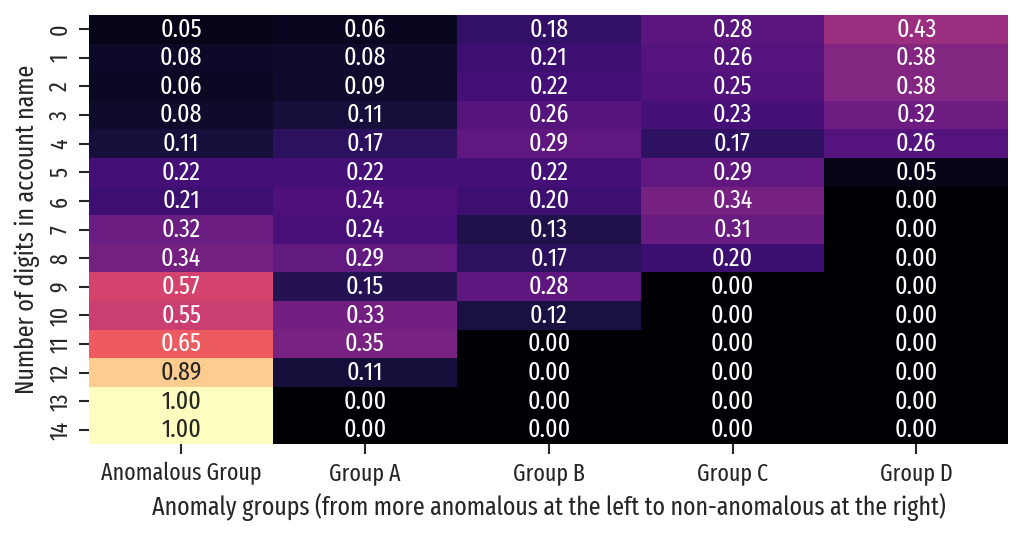}
    \caption{Distribution of accounts according to their anomaly score per number of digits in the username.}
    \label{fig:username_numbers}
\end{figure}

\begin{figure}
    \centering
    \includegraphics[width=0.95\linewidth]{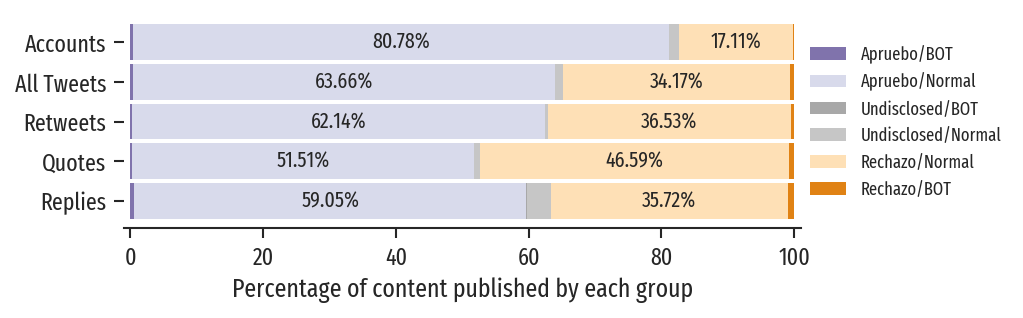}
    \caption{Distribution of content with respect to predicted stance and assigned bot status.}
    \label{fig:content_distribution}
\end{figure}

\begin{figure}
    \centering
    \includegraphics[width=\linewidth]{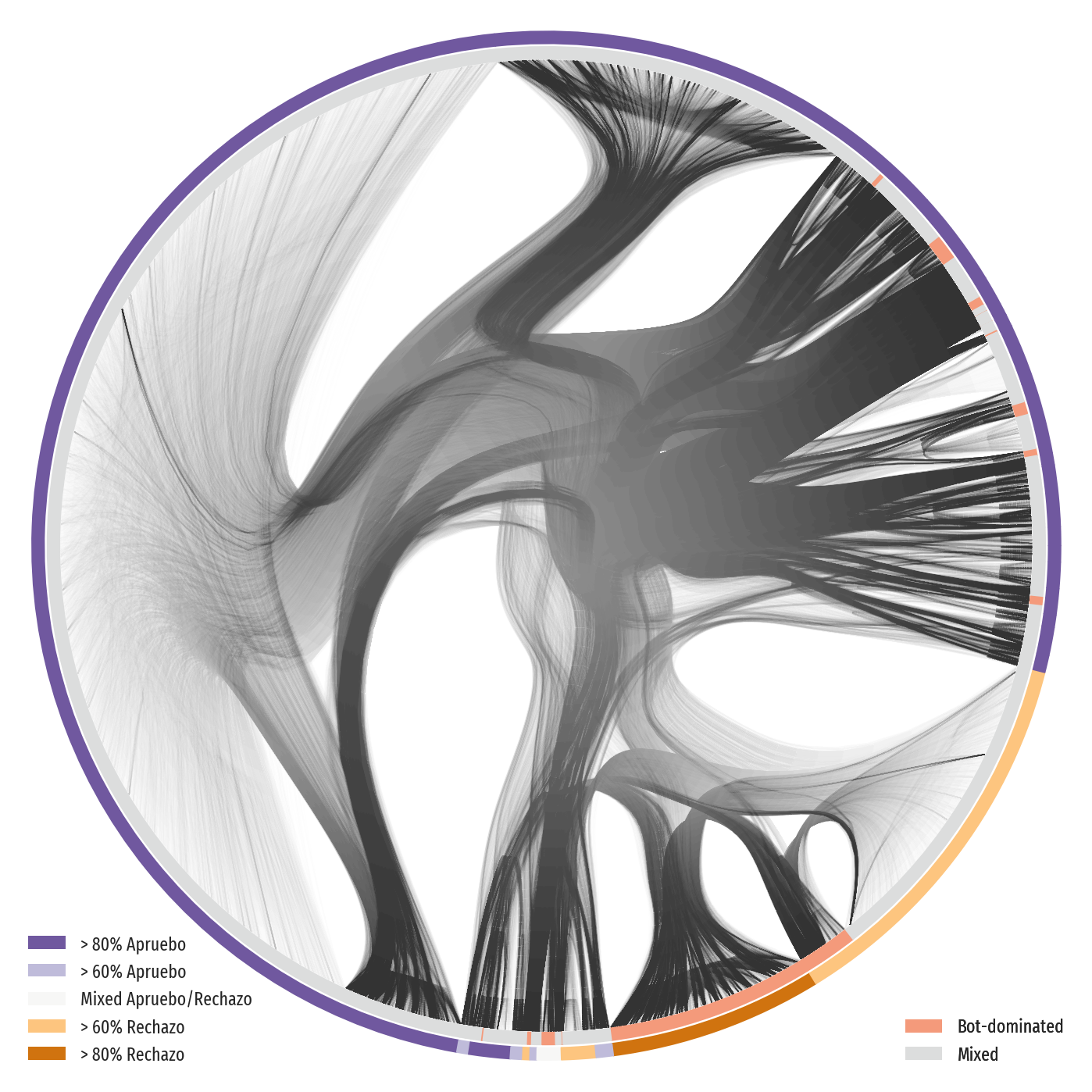}
    \caption{Network of retweets. Communities are represented by the two rings on the outside, one colored according to political stance (outer), and the other colored according to the presence of bots (inner). Edges are lines between nodes, where the origin of the edge (the retweeting account) is colored in light gray, and the destination of the edge (the retweeted account) is colored in dark gray.}
    \label{fig:network}
\end{figure}

Another important feature is the number of digits in a username, as a high number of digits may indicate an account with a randomly generated username. Indeed, the model found that less anomalous accounts tend to have less than four digits in their names (a feasible explanation of this limit is that some accounts have a year in their username), whereas highly anomalous accounts have up to 14 digits (see the distribution in Figure~\ref{fig:username_numbers}).

Then, we assigned bot status to an account that lied in the anomaly group (7.5\% more anomalous), that has been registered from August 8th onward, and that has more than four digits in the username. Only 0.66\% of accounts are labeled as bots with this criteria. Next, we compared the distribution of accounts and the published content taking into account both, predicted stance and bot status (see Figure~\ref{fig:content_distribution}). We observed that \emph{rechazo} publishes a greater amount of content than expected given its number of accounts, however, the activity of bot accounts does not seem suspicious in terms of content volume. This hints that bot activity in this study could be related to bursts of coordinated action, for instance, to establish a trending topic, rather than continuous generation of content. It suggests, as well, that bot detection is tied to political activity, and thus, a generic bot detector may not perform well without considering the political context.

Finally, to explore the potential coordination between bots, we estimated a hierarchical community structure using a Stochastic Block Model~\cite{peixoto2014hierarchical}~(see Figure~\ref{fig:network}), under the assumption that, if there is coordinated action between bots, then those accounts should belong to the same community. We studied the largest connected component of the retweet network, with 143K accounts and 527K weighted edges. In its 118 detected communities, we standardized the fraction of bots within each, and then we classified each into one of three groups: it has more bots than expected (more than one standard deviation of bot presence, 17 communities), and mixed (the rest, 89 communities). We find it interesting that there are no communities without bot accounts, probably indicating that there are false positives in our criteria. There are bot-dominated communities in both political stances, however, the \emph{rechazo} ones are larger. These communities tend to show a high popularity, with inter- and intra-community retweets. This is a sign of coordinated action, although the effects of these actions have yet to be determined.

\section{Conclusions}

In this work we performed a detailed analysis of Twitter discussion in an historical national event in Chile, from the lens of anomalous activity, including bots. We found that, under strict criteria, the number of bots in the discussion is small, and that in terms of produced content, bot accounts do not differ from regular accounts. The difference lies in the network behavior: there are bot-dominated communities in the information diffusion network, and these communities have a high in-degree. This suggests for future work that, although they may not be an army of bots, these small squads in coordination with regular accounts, may influence what is being discussed. This influence has a clear political objective, as \emph{rechazo} (right-leaning) bots form large communities in comparison with \emph{apruebo} bots, which seem to be scattered around larger communities of regular accounts. Since the inner workings of trending topics in Twitter is unknown, this evidence provides information that helps people, journalists and politicians to put the digital discussion into perspective. In particular, bots amplify political polarization and makes more difficult to distinguish the reality with the perception of it.

Future work could focus on strengthening the pipeline of analysis, in particular the bot criteria. On the one hand, previous reports indicate a larger fraction of bots, thus, we may be providing a lower bound of this quantity. On the other hand, our criteria did not include the community detection step, which may help to identify false positives, or the difference between well-behaved and deceiving bots. We will also study how bots behave in the exit referendum of the constitutional process. That is, approving or not the new constitution. 

In terms of representativeness, we acknowledge that Twitter is a biased sample of the population~\cite{baeza2018bias}.
The similarity between our stance prediction and the election results also adds evidence to this aspect. Although~the representativeness of such insights is yet to be~determined, we propose for future work to disentangle national from local representativeness of results.

\begin{acks}
This project uses the graph-tool~\cite{peixoto2014graph}, scikit-learn~\cite{pedregosa2011scikit}, XGBoost~\cite{chen2016xgboost}, numpy~\cite{harris2020array}, pandas~\cite{mckinney2011pandas}, and matplotlib~\cite{hunter2007matplotlib} libraries. 
We thank Paula Vasquez-Henr\'iquez for her comments.
\end{acks}

\newpage
\bibliographystyle{ACM-Reference-Format}
\bibliography{sample-base}

\end{document}